\documentclass[12pt,showpacs,showkeys,amsmath,amssymb]{revtex4}
\usepackage{amsmath,amsfonts,amsthm,amscd,amssymb,latexsym}
\usepackage{bm}
\usepackage{dcolumn}
\usepackage{graphicx}
\usepackage{epstopdf}
\usepackage{color}
\usepackage{epsf}
\usepackage{epsfig}
\usepackage{graphicx, epic, eepic, color}

\begin{document}
\title{Nonlocal Gravity: Modified Poisson's Equation}

\author{C. Chicone}
\email{chiconec@missouri.edu}
\affiliation{Department of Mathematics and Department of Physics and Astronomy, University of Missouri, Columbia,
Missouri 65211, USA }

\author{B. Mashhoon}
\email{mashhoonb@missouri.edu}
\affiliation{Department of Physics and Astronomy,
University of Missouri, Columbia, Missouri 65211, USA}

\begin{abstract} 
The recent nonlocal generalization of Einstein's theory of gravitation reduces in the Newtonian regime to a nonlocal and nonlinear modification of  Poisson's equation of Newtonian gravity. The nonlocally modified Poisson equation implies that nonlocality can simulate dark matter. Observational data regarding dark matter provide limited information about the functional form of the reciprocal kernel, from which the original nonlocal kernel of the theory must be determined. We study this inverse problem of nonlocal gravity in the linear domain, where the applicability of the Fourier transform method is critically examined and the conditions for the existence of the nonlocal kernel are discussed. This approach is illustrated via simple explicit examples for which the kernels are numerically evaluated. We then turn to a general discussion of the modified Poisson equation and present a formal solution of this equation via a successive approximation scheme. The treatment is specialized to the gravitational potential of a point mass, where in the linear regime we recover the Tohline-Kuhn approach to modified gravity. 
\end{abstract}

\pacs{04.20.Cv, 04.50.Kd, 11.10.Lm}

\keywords{nonlocal gravity, nonlocal modification of Newtonian gravity, dark matter}

\maketitle

\section{Introduction}

Newton's inverse-square force law for the universal attraction of gravity successfully explained Kepler's laws of planetary motion and other solar-system observations. However, with the advent of Lorentz invariance, Einstein's general theory of relativity integrated Newtonian gravitation into a consistent relativistic framework that naturally explained the excess perihelion precession of Mercury and has thus far successfully withstood various observational challenges. On the other hand, on galactic scales, the existence of dark matter has been assumed for decades in order to resolve observational problems, such as the ``flat" rotation curves of spiral galaxies. As the nature of dark matter is still unknown, we take the view that what appears as dark matter may very well be a novel aspect of the gravitational interaction. Starting from first principles, a nonlocal generalization of Einstein's theory of gravitation has been developed in recent papers~\cite{nonlocal, NonLocal, BCHM, BM}, where it is demonstrated that nonlocal gravity can naturally simulate dark matter.  
However, the main motivation  for the development of a nonlocal theory of gravitation came from the principle of equivalence, once the general outline of \emph{nonlocal special relativity} had become clear~\cite{BM2}. In nonlocal special relativity, nonlocality is induced by the \emph{acceleration} of observers in Minkowski spacetime. Inertia and gravitation are intimately linked in accordance with the principle of equivalence of inertial and gravitational masses. Hence, gravity is expected to be nonlocal as well. That nonlocality can simulate dark matter emerged later in the course of studying the physical implications of nonlocal general relativity~\cite{nonlocal, NonLocal}.

In this theory of nonlocal gravity, gravitation is described by a local field that satisfies nonlocal integro-differential equations. Thus gravity is in this sense fundamentally nonlocal and its nonlocality is introduced through a ``constitutive" kernel. \emph{This nonlocal kernel is a basic ingredient of the gravitational interaction that must ultimately be determined from observation.} A \emph{direct} nonlocal generalization of the standard form of Einstein's general relativity would encounter severe difficulties~\cite{Bahram:2007}; however, it turns out that one can also arrive at general relativity via a special case of the translational gauge theory of gravity, namely, the  teleparallel equivalent of general relativity. It can be shown that this theory is amenable to generalization through the introduction of a constitutive kernel. In fact, the simplest case of a \emph{scalar} constitutive kernel has been employed in~\cite{nonlocal, NonLocal, BCHM, BM} to develop a consistent nonlocal generalization of Einstein's theory of gravitation. In this approach to nonlocal gravity, nonlocality survives even in the Newtonian regime and appears to provide a natural explanation for ``dark matter"~\cite{SR, C, S, B, G}; that is, nonlocal gravity simulates dark matter. Indeed, the nonlocal theory naturally contains the Tohline-Kuhn phenomenological extension of Newtonian gravity to the galactic realm~\cite{Tohline, Kuhn, Jacob1988}.

Poisson's equation of Newtonian gravity,
\begin{equation}\label{1}
\nabla^2\Phi_{N}(t,\mathbf{x})=4\pi G \rho(t,\mathbf{x})
\end{equation}
is modified in the nonlocal theory to read~\cite{BCHM, BM}
\begin{equation}\label{2}
  \nabla^2\Phi(\mathbf{x})+\sum_i\int\frac{\partial
    \Bbbk(\mathbf{x},\mathbf{y})}{\partial x^i}\frac{\partial\Phi
    (\mathbf{y})}{\partial y^i}d^3y=4\pi G\rho(\mathbf{x})\,.
\end{equation}
Here $\Phi_{N}$ is the Newtonian gravitational potential and  any possible temporal dependence of the gravitational potential $\Phi$ and matter density $\rho$ has been suppressed in Eq.~\eqref{2} for the sake of simplicity. Moreover, the nonlocal kernel $\Bbbk$ is a smooth function of $\mathbf{u}$ and $v$, so that $\Bbbk(\mathbf{x},\mathbf{y}) = K(\mathbf{u},v)$, where
\begin{equation}\label{3}
\mathbf{u} :=  \mathbf{x}-\mathbf{y}, \quad v := \frac{|\nabla_{\mathbf{y}}\Phi(\mathbf{y})|}{|\nabla_{\mathbf{x}}
      \Phi(\mathbf{x})|}.
\end{equation}
Thus $v$ is the source of nonlinearity in Eq.~\eqref{2}. This nonlocal and nonlinear modification of Poisson's equation is invariant under a constant change of the potential, namely, $\Phi \mapsto \Phi + C$, where $C$ is a constant. In addition, Eq.~\eqref{2} satisfies a scaling law; that is, if the matter density is scaled by a \emph{constant} scale factor $s$, $\rho \mapsto s\rho$, then the potential is scaled by the same constant factor, $\Phi \mapsto s \Phi$.

The functional form of the nonlocal kernel is unknown. Let us tentatively assume that it does have a \emph{dominant} linear component $k(\mathbf{u})$ for some gravitational systems; that is, 
\begin{equation}\label{3a}
\Bbbk = k(\mathbf{u})+\kappa(\mathbf{u}, v)\,,
\end{equation}
where $\kappa(\mathbf{u}, v)$ is a relatively small nonlinear perturbation. The physical justification for this supposition is that the implications of the \emph{linear} form of Eq.~\eqref{2}, for the corresponding linear gravitational potential $\Phi_{\ell}$, compare favorably  with the physics of the flat rotation curves of spiral galaxies~\cite{nonlocal, NonLocal, BCHM, BM}. In the linear approximation, the scalar constitutive kernel $\Bbbk$ is only a function of $\mathbf{u}$ and we expect intuitively that even nonlocal gravity would weaken with increasing distance, so that $\Bbbk$ should go to zero as $u:=|\mathbf{u}| \to \infty$. Let us therefore start with Eq.~\eqref{2} and consider, for simplicity, a linear kernel of the form $\Bbbk(\mathbf{x},\mathbf{y}) := k(\mathbf{u})$, so that $\partial \Bbbk / \partial x^i = - \partial \Bbbk / \partial y^i$ in this case. Furthermore, let us assume that as $y:=|\mathbf{y}| \to \infty$, $|\Bbbk(\mathbf{x},\mathbf{y})\nabla_{\mathbf{y}}\Phi_{\ell}(\mathbf{y})|$ falls off to zero faster than $1/y^2$; then, using integration by parts and Gauss's theorem, Eq.~\eqref{2} can be written in this case as 
\begin{equation}\label{2a}
  \nabla^2\Phi_{\ell}(\mathbf{x})+ \int k(\mathbf{x}-\mathbf{y}) \nabla^2\Phi_{\ell}(\mathbf{y})d^3y = 4\pi G\rho(\mathbf{x})\,.
\end{equation}
We assume that this Fredholm integral equation of the second kind has a unique solution~\cite{T}, which can be expressed by means of the reciprocal convolution kernel $q(\mathbf{u})$ as 
\begin{equation}\label{2b}
  \nabla^2\Phi_{\ell}(\mathbf{x}) = 4\pi G\rho(\mathbf{x})+ 4\pi G \int q(\mathbf{x}-\mathbf{y}) \rho(\mathbf{y})d^3y\,.
\end{equation}
The conditions for the validity of this assumption will be examined in the following section. On the other hand, it is an immediate consequence of Eq.~\eqref{2b} that nonlocal gravity acts like dark matter; that is, the gravitational potential in the linear regime is due to the presence of matter of density $\rho$ as well as ``dark matter" of density $\rho_D$ given by
\begin{equation}\label{2c}
\rho_D(\mathbf{x}) =  \int q(\mathbf{x}-\mathbf{y}) \rho(\mathbf{y})d^3y\,.
\end{equation}
Thus the Laplacian of the gravitational potential is given by $4\pi G(\rho+\rho_D)$, where, in the linear approximation, $\rho_D$ is the convolution of $\rho$ and the reciprocal kernel $q$. In particular, for a point mass $M$, $\rho(\mathbf{y})=M\delta(\mathbf{y})$, we have $\rho_D=Mq$.

A similar, but more intricate, result holds when nonlinearity is taken into consideration; that is, nonlocality still simulates dark matter, but the connection between $\rho_D$ and $\rho$ goes beyond Eq.~\eqref{2c}. This assertion is based on the assumption that the nonlocal kernel is of the form of Eq.~\eqref{3a}; then, Eq.~\eqref{2} takes a similar form as Eq.~\eqref{2a}, but with an extra source term ${\cal S}(\mathbf{x})$ due to nonlinearity. That is, 
\begin{equation}\label{2ca}
  \nabla^2\Phi(\mathbf{x})+ \int k(\mathbf{x}-\mathbf{y}) \nabla^2\Phi(\mathbf{y})d^3y = 4\pi G[\rho(\mathbf{x})+ {\cal S}(\mathbf{x})]\,.
\end{equation}
Here ${\cal S}(\mathbf{x})$ can be expressed as the divergence of a vector field, namely, 
\begin{equation}\label{2cb}
{\cal S}=-\nabla \cdot \boldsymbol {\nu}, \quad \boldsymbol {\nu}(\mathbf{x})= \frac{1}{4\pi G}\int \kappa(\mathbf{u}, v)\nabla_{\mathbf{y}}\Phi(\mathbf{y})d^3y\,. 
\end{equation}
Assuming as before that Eq.~\eqref{2ca} has a unique solution via the reciprocal convolution kernel $q(\mathbf{u})$, we find
\begin{equation}\label{2cc}
  \nabla^2\Phi(\mathbf{x}) = 4\pi G\{\rho(\mathbf{x})+ {\cal S}(\mathbf{x}) +  \int q(\mathbf{x}-\mathbf{y}) [\rho(\mathbf{y}) + {\cal S}(\mathbf{y})]d^3y\}\,.
\end{equation}
It follows that the density of  ``dark matter" does have contributions from the nonlinear part of the nonlocal kernel. 

Neglecting nonlinearities, we note that the \emph{linear} relationship between $\Phi_{\ell}$ and $\rho$ implies that one can write
\begin{equation}\label{2d}
\Phi_{\ell}(\mathbf{x}) =  \int {\cal G}(\mathbf{x}-\mathbf{y}) \rho(\mathbf{y})d^3y\,.
\end{equation}
Here the Green function ${\cal G}(\mathbf{u})$ is a solution of
\begin{equation}\label{2e}
 \nabla^2{\cal G}(\mathbf{u}) =  4\pi G [\delta(\mathbf{u})+q(\mathbf{u})]\,.
\end{equation}
If $\rho(\mathbf{x})=M\delta(\mathbf{x})$ for a point mass $M$ in Eq.~\eqref{2b}, we have $\Phi_{\ell}=M{\cal G}$, which means that once the nonlocal gravitational potential is known for a \emph{point mass}, one can determine the potential for any mass distribution via linearity as expressed by Eq.~\eqref{2d}. Assuming that $q(\mathbf{u})$ can be determined from observational data, the inverse problem must be solved to find the linear kernel $k(\mathbf{u})$ from $q(\mathbf{u})$. In this paper, we study this inverse problem in connection with the rotation curves of spiral galaxies; furthermore, we provide a general discussion of the solutions of Eq.~\eqref{2} and investigate some of their physical implications.

Consider the motion of stars within the disk of a spiral galaxy in accordance with the Kepler-Newton-Einstein tradition. For revolution on a circle of radius $r:=|\mathbf{x}|$ about the central spherical  bulge, the speed of rotation $V$ of a star is given by $V^2=G{\cal M}/r$, where ${\cal M}$ is the mass of the bulge, which we take to be the effective mass of the galaxy. Observational data indicate that $V^2$ is nearly constant; therefore, keeping the standard theory forces us to assume that mass ${\cal M}$ has a dark component that increases linearly with $r$. Assuming spherical symmetry for the distribution of this dark matter, we find that
\begin{equation}\label{2f}
\rho_D(\mathbf{x}) =  \frac{V_0^2}{4\pi G} \frac{1}{r^2}\,,
\end{equation}
where $V_0$ is the constant asymptotic velocity of stars in the disk of the spiral galaxy. If dark matter does not exist, but what appears to be dark matter is in fact a manifestation of the nonlocal character of the gravitational interaction, then Eq.~\eqref{2c} together with Eq.~\eqref{2f} implies that $q(\mathbf{u})=u^{-2}/(4\pi \lambda)$ for 
$\rho(\mathbf{x})=M\delta(\mathbf{x})$. Here $M$ is the mass of the point source that represents the spiral galaxy and $\lambda=GM/V_0^2$ is a constant length. With this explicit form for the reciprocal kernel $q$, Eq.~\eqref{2b} becomes identical to an equation that was first introduced by Kuhn in the phenomenological Tohline-Kuhn approach to modified gravity as an alternative to dark matter. 

Let us digress here and briefly mention some relevant aspects of the \emph{linear} phenomenology associated with the flat rotation curves of spiral galaxies. To avoid the necessity of introducing dark matter into astrophysics, Tohline~\cite{Tohline} suggested in 1983 that the Newtonian gravitational potential of a point mass $M$ (representing, in effect, the mass contained in the nuclear bulge of a spiral galaxy) could instead be modified by a logarithmic term of the form
\begin{equation}\label{2g}
\Phi_{\ell}(r)=-\frac{GM}{r}
+\frac{GM}{\lambda}\ln\left(\frac{r}{\lambda}\right) \,.
\end{equation}
Here $GM/ \lambda=V_0^2$, where $V_0$ is the approximately constant rotation velocity of stars in the disk of a spiral galaxy of mass $M$. Thus the constant length $\lambda$ is of the order of 1\,kpc; henceforth, we will assume for the sake of definiteness that $\lambda \approx$ 10 kpc. It follows that in this modified gravity model $M\propto V_0^2$, which disagrees with the empirical Tully-Fisher law~\cite{TullyFisher}. The
Tully-Fisher relation involves a correlation between the infrared luminosity of
a spiral galaxy and the corresponding asymptotic rotation speed $V_0$. This
relation, in combination with other observational data regarding mass-to-light
ratio for spiral galaxies, roughly favors $M\propto V_0^4$, instead of $M\propto V_0^2$
that follows from Tohline's proposal.  On the physical side, however, it should be clear that the Tully-Fisher empirical  relation is based on the electromagnetic radiation emitted by galactic matter and thus contains much more physics than just the law of gravity for a point mass~\cite{Kuhn, Ton}. Tohline's suggestion was taken up and generalized several years later by Kuhn and his collaborators---see~\cite{Kuhn} and an illuminating review of the Tohline-Kuhn work in~\cite{Jacob1988}. Indeed, in Kuhn's linear phenomenological scheme of modified gravity~\cite{Jacob1988}, a nonlocal term is introduced into Poisson's equation, namely, 
\begin{equation}\label{2h}
\nabla^2\Phi_{\ell}=~4 \pi G \Big [\rho+ \frac{1}{4\pi\lambda}
\int \frac{\rho(\mathbf{y})}{|\mathbf{x}-\mathbf{y}|^2}d^3y\Big] \,,
\end{equation}
such that for a point source, $\rho(\mathbf{x})=M\delta(\mathbf{x})$, Eq.~\eqref{2g} is a solution of Eq.~\eqref{2h}. It follows immediately from a comparison of Eq.~\eqref{2b} with Eq.~\eqref{2h} that
\begin{equation}\label{2i}
q(|\mathbf{x}-\mathbf{y}|)=\frac{1}{4\pi\lambda}
\frac{1}{|\mathbf{x}-\mathbf{y}|^2}\,.
\end{equation}
Therefore, to make contact with observational data regarding the rotation curves of spiral galaxies, we suppose, as in previous work~\cite{nonlocal, NonLocal, BCHM, BM}, that the reciprocal kernel $q$ in Eq.~\eqref{2b} is approximately given by the Kuhn kernel in Eq.~\eqref{2i} from the bulge out to the edge of spiral galaxies. 

A remark is in order here regarding the nature of $\lambda$. While for a \emph{point} mass, $\lambda$ in Eq.~\eqref{2g} is a universal constant, the situation may be different for the \emph{interior} potential of a bounded distribution of matter. Consider, for instance, the Newtonian gravitational acceleration for a uniform spherical distribution of density $\rho_0$ and radius $R$. The exterior acceleration has the universal form $d\Phi_{N}/dr = GM/r^2$ for $r>R$, where $M = (4\pi\rho_{0}/3)R^3$; as expected, this is identical to the acceleration for a point source of fixed mass $M$. For $0\le r\le R$, the interior Newtonian gravitational acceleration at a fixed radius $r$ is given by $d\Phi_{N}/dr = (GM/R^3)r$, which decreases with increasing $R$ when $M$ and $r$ are held fixed. Extrapolating from this natural consequence of Newtonian gravity  to the nonlocal domain, one might expect that in the interior of spiral galaxies, for instance, $\lambda$ in Eq.~\eqref{2h} might depend on the size of the system. Indeed, in Kuhn's work, $\lambda$, with a magnitude of more or less around 10 kpc in Eq.~\eqref{2h}, is taken to be larger for larger systems~\cite{Kuhn, Jacob1988}.

Rotation curves of spiral galaxies can thus provide some information regarding the nature of the reciprocal kernel $q(\mathbf{u})$ in the linear case, where Eq.~\eqref{2b} can be directly compared with the Tohline-Kuhn scheme. However, to determine the corresponding nonlocal kernel $k(\mathbf{u})$, we need to know the functional form of $q(\mathbf{u})$ over all space. The simplest possibility would be to assume that $q(u)=u^{-2}/(4\pi \lambda)$ holds over all space; however, the corresponding $k(u)$ does \emph{not} exist---see~\cite{NonLocal}, especially Appendix E, for a detailed discussion of this point. We therefore take up the crucial question of the existence of the linear kernel $k(u)$ for spiral galaxies in sections II and III. The general \emph{nonlinear} inverse problem of nonlocal gravity is beyond the scope of this paper; therefore, we concentrate in sections II and III on the linear problem of finding $k(u)$ from certain very simple extensions of $q(u)$ beyond $q(u)\propto u^{-2}$ that is implied (in the galactic disk) by the flat rotation curves of spiral galaxies.
We then turn to the other main goal of this paper, which is to tackle the general \emph{nonlinear} form of Eq.~\eqref{2}. Unfortunately, the general form of the nonlocal kernel is unknown at present. Nevertheless, a formal treatment of Eq.~\eqref{2} is developed in Sec. IV without recourse to Eq.~\eqref{3a} or any specific assumption about the nature of the nonlocal kernel. As an application of this new approach, we consider the gravitational potential of a \emph{point mass} in Sec. V. Physically, we regard the point mass as an idealization for a spherical distribution of matter; in practice, it will stand, for instance, for the mass of a spiral galaxy, most of which is usually concentrated in a central spherical bulge---see, in this connection, section IV of ~\cite{BCHM}. The \emph{linear} regime, where the kernel is independent of $v$, will be investigated in the first subsection of Sec. V; in fact, we illustrate the effectiveness of our procedure by demonstrating how previous results can be recovered in the new setting. Specifically, we show that the Tohline-Kuhn scheme can be recovered in this case from the general treatment of Sec. IV.  Finally, Sec. VI contains a discussion of our results.

\section{Inverse Problems in Linear Nonlocal Gravity}

The nonlocal theory of gravitation under consideration in this paper is based on the existence of a suitable ``constitutive" kernel $\Bbbk$. For the gravitational potential of spiral galaxies, we assume at the outset that $\Bbbk$ contains a dominant linear part $k(u)$  and a nonlinear part that we can ignore in the context of the present discussion. The reciprocal of $k(u)$ for a spiral galaxy is expected to be of the form of the Kuhn kernel~\eqref{2i} in order that the nonlocal theory could simulate dark matter and be therefore consistent with observational data. It turns out that with the simple form of $q$ given in Eq.~\eqref{2i}, the corresponding reciprocal kernel does \emph{not} exist. That is, if we naturally extend the simple Kuhn kernel to the whole space beyond a galaxy, then there is an infinite amount of simulated dark matter and we have a $q(u)\propto u^{-2}$ for which there is no finite $k(u)$. However, the nonlocal theory is based on the existence of a finite smooth nonlocal kernel. This important problem is taken up in this section.
To determine $k(u)$ from its reciprocal, it is necessary to extend the functional form of kernel~\eqref{2i} so that it becomes smoothly applicable over all space and falls off rapidly to zero at infinity. Let $q(u)$ be this extended reciprocal kernel. We must ensure that its reciprocal $k(u)$, the constitutive kernel of nonlocal gravity in the linear regime, indeed exists and properly falls off to zero as $u\to \infty$. To this end, we show in this section that it is sufficient to require that $q(u)$ and $k(u)$ be smooth absolutely integrable as well as square integrable functions over all space, so that their Fourier integral transforms exist as well. 

In the \emph{linear} Newtonian regime of nonlocal gravity, the nonlocally modified Poisson equation is a  Fredholm integral equation of the second kind, cf. Eq.~\eqref{2a}, 
\begin{equation}\label{30}
g(\mathbf{x})+ \int k(\mathbf{x}-\mathbf{y}) g(\mathbf{y})d^3y = f(\mathbf{x})\ ,
\end{equation}
which we assume has a unique solution that is expressible by means of the reciprocal kernel $q$ as
\begin{equation}\label{31}
f(\mathbf{x})+ \int q(\mathbf{x}-\mathbf{y}) f(\mathbf{y})d^3y = g(\mathbf{x})\ .
\end{equation}
The inverse problem, however, involves finding $k(\mathbf{u})$ once $q(\mathbf{u})$ is known; that is, we wish to obtain Eq.~\eqref{30} starting from Eq.~\eqref{31}. Let us note that our methods can be used in either direction due to the obvious symmetry of Eqs.~\eqref{30} and~\eqref{31}. 

It is useful to express Eq.~\eqref{30} in operator form as $(I+{\cal K})g=f$, where $I$ is the identity operator and $\cal K$ is the convolution operator ${\cal K}g=k\star g $.  Formally, we expect the solution of this equation is  $(I+{\cal Q})f=g$, where ${\cal Q}f=q\star f $. Moreover,  $f=(I+{\cal K})g= (I+{\cal Q})^{-1}g$  would be equivalent to Eq.~\eqref{30}. From a  comparison of Eq.~\eqref{30} with Eq.~\eqref{2a}, we see that the quantities of interest  are  $f=4\pi G \rho$ and $g=  \nabla^2\Phi_{\ell}$.  Here,  the function $\rho$ models the density of matter in space and $\Phi_{\ell}$ is the linear gravitational potential.  Both of these functions can be considered to be   \emph{smooth} in the continuum limit for matter distributions under consideration throughout this work.

\subsection{Liouville-Neumann Method}

It is \emph{formally} possible to obtain Eq.~\eqref{30} from Eq.~\eqref{31}, or the other way around,  by the application of the Liouville-Neumann method of successive substitutions~\cite{T}. That is, we start with Eq.~\eqref{31} and write it in the form
\begin{equation}\label{D1}
f(\mathbf{x}) = g(\mathbf{x})- \int q(\mathbf{x}-\mathbf{y}) f(\mathbf{y})d^3y\ .
\end{equation}
Then, we replace $f(\mathbf{y})$ in the integrand by its value given by Eq.~\eqref{D1}. Iterating this process eventually results in an infinite series---namely, the Neumann series---that may or may not converge. 

A uniformly convergent Neumann series leads to a unique solution of the Fredholm integral equation~\eqref{D1}. Moreover, one can determine $k$ in terms of $q$. The procedure for calculating kernel $k$ of Eq.~\eqref{30} in terms of the iterated $q$ kernels has been discussed, for instance, in~\cite{BCHM}; however, a sign error there in the formula for iterated kernels must be corrected: An overall minus sign is missing on the right-hand side of Eq. (3) of~\cite{BCHM}. The spherical symmetry of $q$, in the cases under consideration in this section, implies that all iterated kernels are functions of $u$. Thus let $q_n(u)$, $n=1,2,...$, be the relevant iterated kernels such that $q_1=q$ and 
\begin{equation}\label{D2}
q_{n+1}(|\mathbf{x}-\mathbf{y}|) = - \int q(|\mathbf{x}-\mathbf{z}|) q_n(|\mathbf{z}-\mathbf{y}|)d^3z\ .
\end{equation}
It follows that in this case the nonlocal kernel is
\begin{equation}\label{D3}
k(|\mathbf{x}-\mathbf{y}|) = -\sum_{n=1}^{\infty} q_n(|\mathbf{x}-\mathbf{y}|)\ .
\end{equation}

In trying to determine $k$ from $q$, one can therefore start from the study of the Neumann series.
Using the approach developed in~\cite{T},  and working in the Hilbert space of square-integrable functions, it can be shown by means of the Schwarz inequality that the Neumann series converges and the nonlocal kernel exists for $\lVert q \rVert <1$; moreover, the solution of the Fredholm integral equation~\eqref{31} by means of the Neumann series is \emph{unique}. However, the norms of the convolution operators of interest in our work are not square integrable, since
\begin{equation}\label{32}
\lVert q \rVert^2=  \int q^{2}(\mathbf{x}-\mathbf{y}) d^3x d^3y = \int d^3x \int q^2(\mathbf{u})d^3u=\infty \,,
\end{equation}
so that the standard Hilbert space theory developed in~\cite{T} is not applicable here. That is, the $L^2$ norm of $q$ could be finite, but the norm of the corresponding convolution operator in $L^2$ is infinite.

On the other hand, let us suppose that the space of functions of interest is a Banach space ${\cal B}$ and that ${\cal Q}$ is a bounded linear operator on ${\cal B}$, ${\cal Q}: {\cal B} \to {\cal B}$,  with $\lVert {\cal Q} \rVert<1$; then, one can show that $(I+{\cal Q})$ has an inverse given by $\sum_{n=0}^{\infty}(-{\cal Q})^n$, where the series converges uniformly in the set of all bounded linear operators on ${\cal B}$. In this formula, $(-{\cal Q})^n$ in the series corresponds to the iterated kernel $q_n$, $n=1,2,3,...,$ in the Neumann series and the convergence of the series is equivalent to the existence of  kernel $k$. It seems that the sufficient condition $\lVert {\cal Q} \rVert<1$ for the convergence of the Neumann series cannot be satisfied for any physically reasonable extension of the Kuhn kernel~\eqref{2i}; that is, our various attempts in this direction have been unsuccessful. In short, the Neumann series diverges; therefore, we resort to the Fourier transform method in this work. 

We caution that our mathematical approach may not be unique, as the theory may work in other function spaces that we have not considered here. The general mathematical problem is, of course, beyond the scope of this paper.

\subsection{Fourier Transform Method}

Following well-known mathematics (see, for example,~\cite[\S 9.6] {PS}), precise conditions can be determined for our convolution operators to be invertible. The basic idea is to use the Fourier transform $\mathcal{F}$ defined for $L^1$ functions (that is,  functions which are absolutely integrable over all of space) by
\begin{equation}\label{33}
\mathcal{F}[f](\boldsymbol{\xi}) =\hat{f} (\boldsymbol{\xi}) =  \int f(\mathbf{x}) e^{-i \boldsymbol{\xi} \cdot \mathbf{x}}~ d^3x\,
\end{equation}
to prove a basic lemma:  \emph{ If $q$ is in $L^1$,  then its Fourier transform $\hat q$ is continuous and the convolution operator ${\cal Q}$ given by ${\cal Q}f=q \star f$  is a bounded operator on $L^2$ whose spectrum is the closure of the range of the $\hat q$.}
The proof outlined here uses  standard results of $L^2$ theory. By an elementary argument, the Fourier transform of an arbitrary $L^1$ function is continuous. A deeper result (Plancherel's theorem) states that the Fourier transform can be extended to a bounded invertible operator on $L^2$ that preserves the $L^2$ norm; in other words,  the extended Fourier transformation is an isometric isomorphism of the Hilbert space $L^2$.  This extended operator (also denoted by $ \mathcal{F}$ and simply called the Fourier tranform) maps convolutions to products: ${\cal F}[f\star g]={\cal F}[f] {\cal F}[g]$.  Let  $\hat{\mathcal{Q}}$ be the 
(multiplication) operator  defined on $L^2$ by $\hat{\mathcal{Q}} f=\hat{q} f$, where $\hat q$ is the Fourier transform of the kernel of $\cal Q$.   If follows from the definitions and the action of the Fourier transform on convolutions that $\mathcal{Q}= {\cal F}^{-1} \hat{ \mathcal{Q}} {\cal F}$; therefore, the spectra of  the operators $\mathcal{Q}$  and $\hat{\mathcal{Q}}$ coincide. The spectrum of the multiplication operator  $\hat{ \mathcal{Q}}$  (with its continuous multiplier $\hat q$) on $L^2$ is the closure of the range of $\hat{q}$, as required. 

A corollary of the  lemma  is  the result that we require to analyze our integral equations: \emph{If  the number $-1$ is not in the closure of the range of $\hat q$, then  $I+\mathcal{Q}$ is invertible. }

To see how these ideas can be used to obtain Eq.~\eqref{30} more explicitly, we consider  integral equation~\eqref{31} in the form 
$f+q\star f=g$ and apply the Fourier transform---under the assumption that $f$ and $g$ are in $L^2$ and $q$ is in $L^1$---to obtain the equivalent equation 
\begin{equation}\label{34}
(1+\hat{q})\hat{f} =  \hat{g}\,.
\end{equation}
If $1+\hat{q}\ne 0$, then 
\begin{equation}\label{35}
 \hat{f}=\frac{1}{1+\hat{q}} \hat {g}= (1+\frac{-\hat{q}}{1+\hat{q}})  \hat {g}= \hat {g}+\tilde{ k} \hat {g}\,,
 \end{equation}
 where 
  \begin{equation}\label{35.1}
\tilde k:=\frac{-\hat{q}}{1+\hat{q}}\,.
 \end{equation}
 Applying the inverse Fourier transform, it follows that 
 \begin{equation}\label{36}
f= g+ {\cal F}^{-1}[\tilde k \hat g]\,.
 \end{equation}
\emph{ If there is an $L^1$  function $k$ such that  ${\cal F}[k]=\tilde k$, then}
\begin{equation}\label{38}
f=g+k\star g 
\end{equation}
\emph{and $I+\mathcal{K}$,  where  $\mathcal{K} g=k\star g$, is the inverse of $I+\mathcal{Q}$.}

These results are illustrated in detail in the next section via  specific examples.  In these applications, we will employ a useful lemma regarding the Fourier sine transform.  Consider the integral
\begin{equation}\label{40}
J(\xi)= \int_0^{\infty} h(x, \xi)\sin(\xi x)dx\,.
\end{equation}
\emph{For each $\xi \in (0, \infty)$, let $h(x,\xi)$  be a smooth positive integrable function that monotonically decreases over the interval of integration; then, $J(\xi)>0$.} To prove this assertion for each $\xi >0$, we divide the integration interval in Eq.~\eqref{40} into segments $(2\pi \xi^{-1}n, 2\pi \xi^{-1}n +2\pi \xi^{-1})$ for $n = 0, 1, 2, ...$. In each such segment, the corresponding sine function, $\sin(\xi x)$, goes through a complete cycle and is positive in the first half and negative in the second half. On the other hand, the monotonically decreasing function $h(x, \xi)>0$ is consistently larger in the first half of the cycle than in the second half; therefore, the result of the integration over each full cycle is positive and consequently $J(\xi)>0$.

For $\xi \to 0$, $\sin (\xi x) \to 0$ and hence $J(0)=0$, while for $\xi \to \infty$, the integration segments shrink to zero and $J$ tends to $0$ in the limit as $\xi \to \infty$, if the corresponding limit of $h(x, \xi)$ is finite everywhere over the integration domain. This latter conclusion is, of course, a variation on the Riemann-Lebesgue lemma.

\section{Existence of the Linear Kernel: Examples}

We wish to determine $k(u)$ from a knowledge of $q(u)$. Let us note that if $q(\mathbf{u})$  is only a function of the radial coordinate $u$, Eq.~\eqref{33} reduces to 
\begin{equation}\label{B2}
\hat{q}(\xi)=\frac{4\pi}{\xi}\int_0^{\infty}rq(r)\sin(\xi r)dr\,, 
\end{equation}
where $\xi:0\to \infty$ is the magnitude of $\boldsymbol{\xi}$. That is, in Eq.~\eqref{33}, we introduce spherical polar coordinates $(r, \theta, \phi)$ and imagine that the coordinate system is so oriented that $\boldsymbol{\xi}$ points along the polar axis; then, the angular integrations can be simply carried out using the fact that 
\begin{equation}\label{B3}
\int_0^{\pi}e^{-i \xi r \cos \theta}\sin \theta d\theta=2 \frac{\sin(\xi r)}{\xi r}\,. 
\end{equation}
In general, the Fourier transform  $\hat{q}(\boldsymbol{\xi})$ of a square-integrable function  $q(\mathbf{u})$ is square integrable. In case  $1+\hat{q}\ne 0$ and $\hat{k}(\boldsymbol{\xi})$ is an $L^2$ function, the Fourier transform method of the previous section becomes applicable here. Thus a suitable nonlocal kernel $k(\mathbf{x})$ exists in this case and is given by
\begin{equation}\label{B4}
k(\mathbf{x}) = \frac{1}{(2\pi)^3} \int \hat{k} (\boldsymbol{\xi}) e^{i \boldsymbol{\xi} \cdot \mathbf{x}}~ d^3\boldsymbol{\xi}\,, \quad \hat{k} (\boldsymbol{\xi})=-\frac{\hat{q} (\boldsymbol{\xi})}{1+\hat{q} (\boldsymbol{\xi})}\,.
\end{equation}
For the explicit radial examples under consideration, Eq.~\eqref{B4} takes the form
\begin{equation}\label{B5}
k(r)=-\frac{1}{2\pi^2 r}\int_0^{\infty}\frac{\xi\hat{q}(\xi)}{1+\hat{q}(\xi)}\sin(\xi r)d\xi\,.
\end{equation}

\subsection{An Example}

To extend Kuhn's kernel, Eq.~\eqref{2i}, smoothly over all space, one may consider, for instance, 
\begin{equation}\label{BB1}
q(u)=\frac{1}{4\pi \lambda}\frac{d}{du}\Big[-\frac{F(u)}{a+u}\Big]\,.
\end{equation}
Here $a$ is a constant length scale characteristic of the radius of nuclei of spiral galaxies, while $F(u)$ is a smooth function that is nearly unity over much of the interval $(0, A)$, but then rapidly drops off to zero as $u\to \infty$. The constant $A$ represents another length scale characteristic of the radius of galactic disks. Thus $a\ll \lambda < A$ under physically reasonable conditions, and for $a\ll u < A$, one can show that Eq.~\eqref{BB1} essentially coincides with Kuhn's kernel. To recover the flat rotation curves of spiral galaxies, Eq.~\eqref{BB1} should agree with Kuhn's kernel from the bulge, which is, say, at a distance of about $\lambda \approx 10$ kpc from the galactic center, out to the edge of the galaxy, which is, say, at a distance of about $3\lambda$. The function $F(u)$ can be chosen so as to render $1+\hat{q} > 0$; to illustrate this point, we choose
\begin{equation}\label{BB2}
F(u)= e^{-u/A}\,.
\end{equation}

Let us now consider the application of Eqs.~\eqref{B2} and~\eqref{B5} to the particular case of Eqs.~\eqref{BB1} -- \eqref{BB2}. Regarding the parameters that appear in these equations, it is sufficient to suppose at the outset only that $\lambda>0$, $a>0$ and $A>0$ are length scales of interest; moreover, we introduce, for the sake of simplicity,
\begin{equation}\label{B6}
\alpha :=\frac{1}{A}\,.
\end{equation}
We note that $q(u)$ is smooth and positive everywhere and rapidly decreases to zero at infinity; indeed, $q(u)$ is integrable as well as square integrable. It is preferable to work with dimensionless quantities; thus, we let all lengths---such as $r$, $u$, $\lambda$, $a$ and $A$---be expressed in units of $\lambda$. Then, $\lambda^3q$, $\lambda \xi$, $\lambda \alpha$ and $\hat{q}$ are dimensionless. Similarly, $\lambda^3k$ and $\hat{k}$ are dimensionless. Henceforth, we deal with these dimensionless quantities; in effect, this means that $\lambda=1$ in the following formulas. \emph{It is possible to show that in this case for any $\xi \ge 0$, $\hat{q}(\xi)> -a$.}

Substituting Eqs.~\eqref{BB1} and~\eqref{BB2} in Eq.~\eqref{B2} and integrating by parts, we find
\begin{equation}\label{B7}
\hat{q}(\xi)=\frac{1}{\xi}\int_0^{\infty}\frac{e^{-\alpha r}}{a+r}~\frac{d}{dr}[r\sin(\xi r)]~dr\,. 
\end{equation}
Next, differentiating $r\sin(\xi r)$ and noting that 
\begin{equation}\label{B8}
\sin(\xi r) + \xi r \cos(\xi r)=[\sin(\xi r)-a \xi \cos(\xi r)]+(a+r)\xi \cos(\xi r)\,,
\end{equation}
Eq.~\eqref{B7} can be written as
\begin{equation}\label{B9}
\hat{q}(\xi)= {\cal I} + \int_0^{\infty}e^{-\alpha r}\cos(\xi r)dr\,, 
\end{equation}
where
\begin{equation}\label{B10}
\int_0^{\infty}e^{-\alpha r}\cos(\xi r)dr =\frac{\alpha}{\alpha^2 + \xi^2}\,, 
\end{equation}
according to formulas 3.893 on page 477 of Ref.~\cite{G+R}, and
 \begin{equation}\label{B11}
{\cal I}=\frac{1}{\xi}\int_0^{\infty}\frac{e^{-\alpha r}}{a+r}~[\sin(\xi r)-a \xi \cos(\xi r)]~dr\,. 
\end{equation}
Let us now introduce an angle $\gamma$ connected with $a\xi$ such that
\begin{equation}\label{B12}
a\xi :=\tan \gamma\,,
\end{equation}
and note that as $\xi: 0\to\infty$, we have $\gamma: 0\to \pi/2$ and $\gamma/\xi: a\to 0$. It is useful to  introduce a new variable $X$ in Eq.~\eqref{B11}, $r=X+\gamma/\xi$, since 
\begin{equation}\label{B13}
\sin(\xi r)-a \xi \cos(\xi r)=\frac{1}{\cos \gamma}\sin(\xi r - \gamma)\,.
\end{equation}
Then, Eq.~\eqref{B11} can be written as 
\begin{equation}\label{B14}
{\cal I}=\frac{e^{-\alpha \gamma/\xi}}{\xi \cos \gamma}\int_{-\gamma/\xi}^{\infty}~\frac{e^{-\alpha X}}{a+\frac{\gamma}{\xi}+X}~\sin(\xi X)~dX\,. 
\end{equation}
In this expression, the integration from $X=-\gamma/\xi$ to $\infty$ can be expressed as a sum of two terms, one from $0$ to $\infty$ and the other from $X=-\gamma/\xi$ to $0$. That is, ${\cal I}={\cal P}+{\cal N}$, where
\begin{equation}\label{B15}
{\cal P}=\frac{e^{-\alpha \gamma/\xi}}{\xi \cos \gamma}\int_0^{\infty}~\frac{e^{-\alpha X}}{a+\frac{\gamma}{\xi}+X}~\sin(\xi X)~dX 
\end{equation}
is positive by the argument presented at the end of Sec. II, since $\exp(-\alpha X)/(X+a+\gamma/\xi)$ is a smooth positive integrable function that monotonically decreases for $X: 0\to \infty$, and 
\begin{equation}\label{B16}
{\cal N}=\frac{e^{-\alpha \gamma/\xi}}{\xi \cos \gamma}\int_{-\gamma/\xi}^0~\frac{e^{-\alpha X}}{a+\frac{\gamma}{\xi}+X}~\sin(\xi X)~dX\,, 
\end{equation}
which is negative. This latter point can be made explicit by introducing a new variable $\xi X = - Y$ into Eq.~\eqref{B16}; then, we have
\begin{equation}\label{B17}
-\xi \cos \gamma ~ {\cal N}(\xi)=\int_0^{\gamma}~\frac{e^{-\frac{\alpha}{\xi}(\gamma-Y)}}{(\gamma-Y)+a\xi}~\sin Y~dY\,. 
\end{equation}
The right-hand side of this equation involves an integrand that increases monotonically from $0$ to $\sin \gamma/(a\xi)$  as $Y: 0\to \gamma$. Thus the right-hand side of Eq.~\eqref{B17} is less than $\gamma \sin \gamma/(a\xi)$; consequently, ${\cal N}(\xi)> -\gamma / \xi$ by Eq.~\eqref{B12}. As $0\le \gamma / \xi \le a$, we conclude that ${\cal N}(\xi)> -a$. Collecting our results, we therefore have that $\hat{q}> - a$ and 
\begin{equation}\label{B18}
1+\hat{q}(\xi) > 1-a\,.
\end{equation}
\emph{Hence the Fourier transform method of the previous section is applicable to Eq.~\eqref{BB1} if $a<1$. It then follows from Eq.~\eqref{B4} that $|\hat{k}|<|\hat{q}|/(1-a)$, so that $\hat{k}(\xi)$ is in $L^2$ as well, and we can find the nonlocal kernel $k$ from Eq.~\eqref{B5}.} 

In connection with the rotation curves of spiral galaxies, it is useful to consider the amount of sham dark matter that is associated with such a model. An estimate of the net amount of simulated dark matter $M_D$ can be obtained from the integration of Eq.~\eqref{2c} over all space, where we set 
$\rho(\mathbf{y})=M\delta(\mathbf{y})$ for the sake of simplicity. Thus
\begin{equation}\label{B19}
\frac{M_D}{M} \approx 4\pi \int_0^\infty r^2q(r)dr\,.
\end{equation}
For the example under consideration, we find from Eqs.~\eqref{BB1},~\eqref{BB2} and~\eqref{B19} via integration by parts that 
\begin{equation}\label{B20}
\frac{M_D}{M} \approx \frac{2}{\alpha} \int_0^\infty\frac{\zeta e^{-\zeta}}{a\alpha + \zeta}d\zeta\,,
\end{equation}
where the definite integral can be expressed in terms of the exponential integral function as $1+a\alpha \exp(a\alpha) {\rm Ei} (- a\alpha)$---see, for instance, page 311 of Ref.~\cite{G+R}. Thus nonlocality in this case simulates, in effect, a net amount of dark matter that is nearly $2/\alpha$ times the actual mass $M$ of the galaxy, since the integral term on the right-hand side of Eq.~\eqref{B20} is nearly unity for physically reasonable values of the parameters, namely, $0<a\alpha \ll 1$. Indeed, for $0<x\ll 1$, ${\rm Ei}(-x)\approx {\cal C} + \ln x$, where ${\cal C}=0.577...$ is the Euler-Mascheroni constant---see page 927 of Ref.~\cite{G+R}. Further considerations involving kernels $q$ and $k$ are relegated to Appendix A. 

Choosing dimensionless parameters $\alpha= 0.1$ and $a= 0.001$ in this example, the corresponding numerical results for $\hat{q}$ and $k$ are presented in Figures 1 and 2.

\subsection{A Second Example}

The purpose of this subsection is to discuss a second example of a nonlocal kernel. We start with the reciprocal kernel
\begin{equation}\label{C1}
q(u)=\frac{1}{4\pi \lambda}~ \frac{1+\alpha(a+u)}{u(a+u)}~e^{-\alpha u}\,.
\end{equation}
Here $\alpha :=1/A$, as before. This reciprocal kernel has a \emph{central cusp} and behaves as $1/u$ for $u\ll a$, which is reminiscent of the density of dark matter in certain dark matter models---see, for instance, Ref.~\cite{S}. Moreover, for $a \ll u<A$, $q$ behaves like the Kuhn kernel, while for $u\gg A$, it falls off exponentially to zero.

Kernel~\eqref{C1} is a smooth positive integrable function that is in $L^2$. Using dimensionless quantities, Eq.~\eqref{B2} takes the form
\begin{equation}\label{C2}
\hat{q}(\xi)=\frac{1}{\xi}\int_0^{\infty}(\alpha + \frac{1}{a+r})e^{-\alpha r}\sin(\xi r)dr\,. 
\end{equation}
From formulas 3.893 on page 477 of Ref.~\cite{G+R}, we find
\begin{equation}\label{C3}
\hat{q}(\xi)=\frac{\alpha}{\alpha^2 + \xi^2} + \frac{1}{\xi}\int_0^{\infty}\frac{e^{-\alpha r}}{a+r}\sin(\xi r)dr\,. 
\end{equation}
Here we can directly use the lemma given at the end of Sec. II, since $\exp(-\alpha r)/(a+r)$ is a smooth positive integrable function that decreases monotonically for $r: 0 \to \infty$,  to conclude that for $0\le \xi <\infty$, $\hat{q}(\xi)> 0$, while $\hat{q}(\xi)\to 0$ as $\xi \to \infty$ by the Riemann-Lebesgue lemma. It follows that $|\hat{k}(\xi)| \le  |\hat{q}(\xi)|$, so that $\hat{k}$ is in $L^2$ as well and the nonlocal kernel $k$ can be determined via Eq.~\eqref{B5}.

\begin{figure}\label{Fig:1} 
\includegraphics[scale=0.8,angle=0]{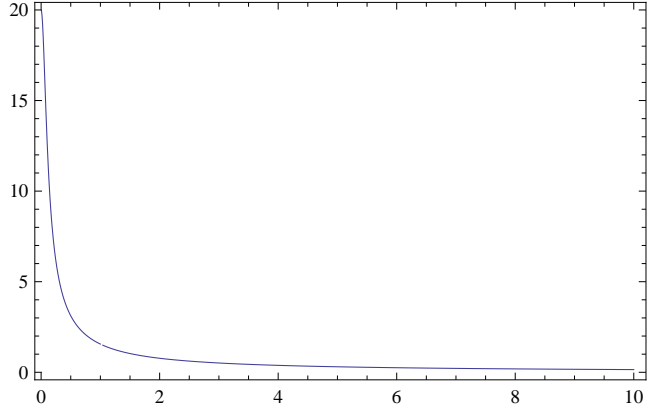}
\caption{Plot of $\hat{q}$ versus $\lambda \xi$ for the reciprocal kernel $q$ given in Eqs.~\eqref{BB1} and~\eqref{BB2}. The parameter values are $\lambda \alpha = 0.1$ and $a/\lambda = 0.001$. The function $\hat{q}$ starts from $\hat{q}(0) \approx 20$ and rapidly falls off initially, but then slowly decreases to zero as $\lambda \xi \to \infty$.} 
\end{figure}

\begin{figure}\label{Fig:2} 
\includegraphics[scale=0.8,angle=0]{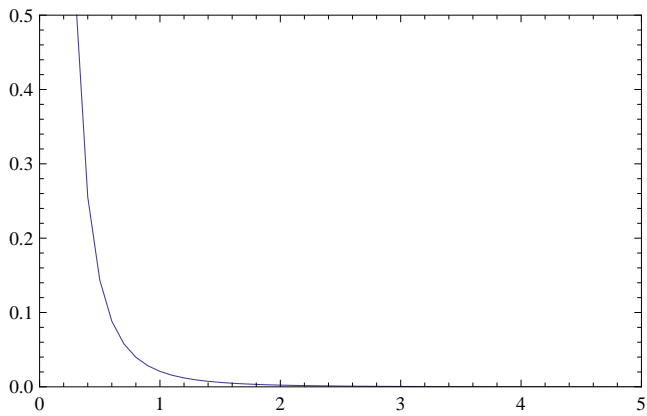}
\caption{Plot of $-\lambda^3 k$ versus $u/\lambda$ when the reciprocal kernel $q$ is given by Eqs.~\eqref{BB1} and~\eqref{BB2}. The parameter values are $\lambda \alpha = 0.1$ and $a/\lambda = 0.001$, just as in Figure 1. The function $-\lambda^3 k$ starts from $\approx 80101$ at $u=0$ and drops off to nearly zero very fast; in fact, for $u/\lambda \ge 2.5$ it is essentially zero at the level of accuracy of this plot.} 
\end{figure}

\begin{figure}\label{Fig:3} 
\includegraphics[scale=0.8,angle=0]{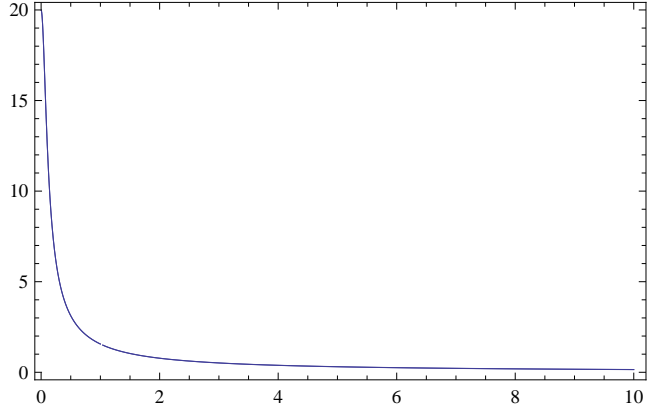}
\caption{Plot of $\hat{q}$ versus $\lambda \xi$ for the reciprocal kernel $q$ given in Eq.~\eqref{C1}. The parameter values are $\lambda \alpha = 0.1$ and $a/\lambda = 0.001$, just as in Figure 1. As pointed out in Appendix A, $\hat{q}$ is in this case always slightly larger than the one in Figure 1, but this is hardly noticeable for the parameter values under consideration. For instance, at $\lambda \xi =2$, $\hat{q} \approx 0.773$ in Figure 1, while here $\hat{q} \approx 0.779$.} 
\end{figure}

\begin{figure}\label{Fig:4} 
\includegraphics[scale=0.8,angle=0]{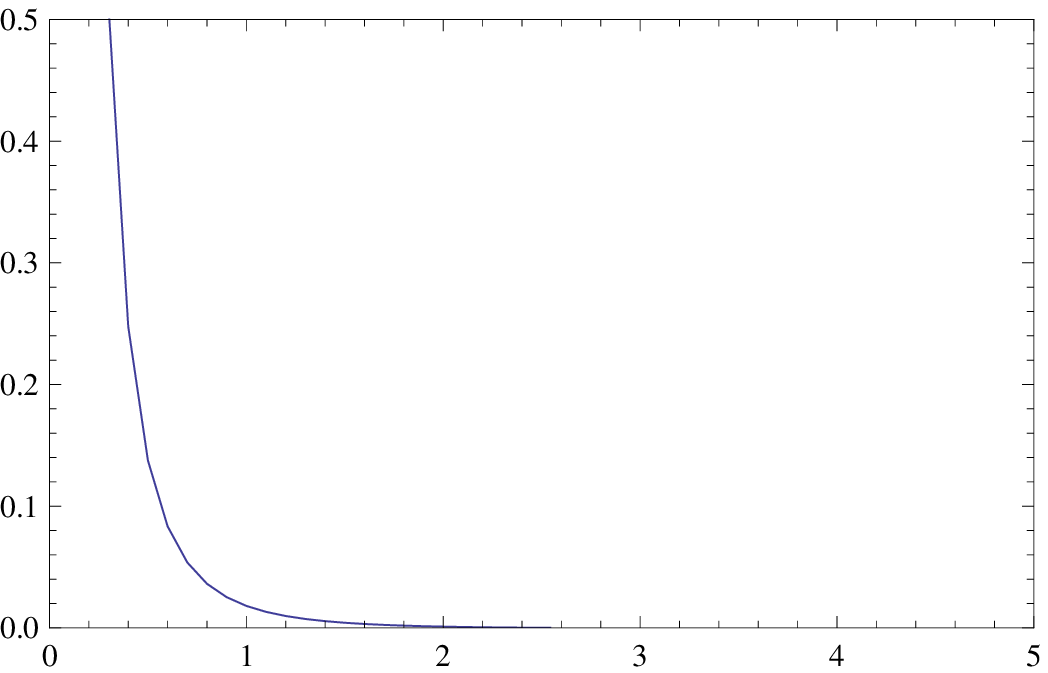}
\caption{Plot of $-\lambda^3 k$ versus $u/\lambda$ when the reciprocal kernel $q$ is given by Eq.~\eqref{C1}. The parameter values are $\lambda \alpha = 0.1$ and $a/\lambda = 0.001$, just as in Figure 3. The function $-\lambda^3 k$ starts from  $\infty$ at $u=0$ and drops off to nearly zero very fast; in fact, for $u/\lambda \ge 2.5$ it is essentially zero at the level of accuracy of this plot. Though this figure appears to be indistinguishable from Figure 2 in the plotted range, their numerical values are indeed different.} 
\end{figure}

In this case, the analog of Eq.~\eqref{B19} is given by
\begin{equation}\label{C4}
\frac{M_D}{M} \approx \frac{2}{\alpha} + a e^{a\alpha} {\rm Ei}(-a\alpha)\,,
\end{equation}
which is, for $0<a\alpha \ll 1$, nearly the same as in Eq.~\eqref{B20}. For this second example, the numerical results involving $\hat{q}$ and $k$ for $\alpha= 0.1$ and $a= 0.001$ are presented in Figures 3 and 4. 

Appendix A contains further useful mathematical results relevant to the examples described in this section and the corresponding numerical work presented in the figures.

The nonlocal ``constitutive'' kernel $k(u)$ turns out to be negative for models of spiral galaxies under consideration in this work. Moreover, as Figures 2 and 4 indicate, nonlocality in this case involves sampling sufficiently close spatial regions. Indeed, around any point $\mathbf{x}$, the influence of the field amplitude at point $\mathbf{y}$ may be significant only when $u=|\mathbf{x}-\mathbf{y}|$ is smaller than around $2.5\lambda$, or about 25 kpc. \emph{In fact, as expected, at any given field point, the nonlocal influence of the field amplitude at a nearby point decreases with increasing distance extremely fast.}

\section{Nonlocal and Nonlinear Poisson's Equation}

The purpose of this section is to present the main outlines of a formal approach to the modified Poisson equation with a general nonlinear kernel. That is, we do \emph{not} assume here that the nonlocal kernel consists of a dominant linear part together with a small nonlinear perturbation.

The right-hand side of Eq.~\eqref{2} can be replaced by the Laplacian of the Newtonian potential via Eq.~\eqref{1}. Furthermore, it is straightforward to see that the nonlocal contribution to Eq.~\eqref{2} can be written as the divergence of a vector field. It follows from these remarks that  modified Poisson's equation can thus be written as $\nabla \cdot \boldsymbol{\Psi}=0$, where 
\begin{equation}\label{4}
\boldsymbol{\Psi} =  \nabla \Phi + \int \Bbbk(\mathbf{x},\mathbf{y})\nabla_{\mathbf{y}}\Phi(\mathbf{y})d^3y - \nabla \Phi_{N}\,.
\end{equation}
For a bounded matter distribution, we can write the solution of Eq.~\eqref{1} as
\begin{equation}\label{4a}
 \Phi_{N}(\mathbf{x}) = -G \int \frac{\rho(\mathbf{y})}{|\mathbf{x} - \mathbf{y}|}d^3y,
\end{equation}
so that, as is customary, the Newtonian gravitational potential is assumed to be zero at infinity~\cite{Kellogg}. 

We are interested in the solution of the nonlinear integral equation
\begin{equation}\label{5}
 \nabla \Phi + \int \Bbbk(\mathbf{x},\mathbf{y})\nabla_{\mathbf{y}}\Phi(\mathbf{y})d^3y = \nabla \Phi_{N} + \boldsymbol{\Psi}\,.
\end{equation}
The divergence-free vector field $\boldsymbol{\Psi}$ must be such that Eq.~\eqref{5} is integrable. Indeed, the integrability condition for Eq.~\eqref{5} is that 
\begin{equation}\label{6}
 \int \nabla_{\mathbf{x}} \Bbbk(\mathbf{x},\mathbf{y}) \times \nabla_{\mathbf{y}}\Phi(\mathbf{y})d^3y =  \nabla \times \boldsymbol{\Psi}\,.
\end{equation}
In effect, the modified Poisson equation has thus been once integrated and reduced to Eqs.~\eqref{5} and \eqref{6}, which are, however, still unwieldy.

Let $\mathbf{U}$ represent the left-hand side of Eq.~\eqref{6}; then, one can express Eq.~\eqref{6} as $\nabla \times \boldsymbol{\Psi}=\mathbf{U}$, where $\mathbf{U}$ is divergence-free. The divergence of $\boldsymbol{\Psi}$ vanishes and its curl is $\mathbf{U}$; therefore, the curl of $-\mathbf{U}$ is the Laplacian of $\boldsymbol{\Psi}$,
\begin{equation}\label{7}
\nabla^2\boldsymbol{\Psi}=-\nabla \times \mathbf{U}\,.
\end{equation}
If $\mathbf{U}(\mathbf{x})$ is bounded for small $r=|\mathbf{x}|$, falls off to zero faster than $1/r$ for large $r$ and $\boldsymbol{\Psi} \to 0$ as $r \to \infty$, then
\begin{equation}\label{8}
\boldsymbol{\Psi}(\mathbf{x})=\frac{1}{4\pi} \nabla \times \int \frac{\mathbf{U}(\mathbf{y})}
{|\mathbf{x}-\mathbf{y}|}d^3y\,,  \quad \mathbf{U}(\mathbf{x})=\nabla \times \int \Bbbk(\mathbf{x},\mathbf{y})\nabla_{\mathbf{y}}\Phi(\mathbf{y})d^3y\,. 
\end{equation}
In this way, one can determine $\boldsymbol{\Psi}$ from the integrability condition, namely, Eq.~\eqref{6}, and substitute it back into Eq.~\eqref{5}.

\subsection{Formal Solution via Successive Approximations}

The solution of the integral equation for the gravitational potential $\Phi$ is expected to consist of the Newtonian gravitational potential $\Phi_N$ together with nonlocal corrections as in a Neumann series; therefore, it is natural to devise a formal solution of Eq.~\eqref{5} using the method of successive approximations~\cite{T}. In view of the divergence of the Neumann series in connection with the considerations of the first part of this paper, we must assume here that the gravitational system under consideration in this subsection cannot be approximated by a point mass. Let $\Phi_0, \Phi_1, ..., \Phi_n, ...$ be a series of approximations to the gravitational potential such that $\Phi_0:=\Phi_N$ and $\Phi_n$ approaches $\Phi$ in the limit as $n \to \infty$. Moreover, for $n>0$, we define $\Phi_n$ to be such that 
\begin{equation}\label{9}
 \nabla \Phi_1 =  \nabla \Phi_N +  \boldsymbol{\Psi}_0 - \int K(\mathbf{u}, v_0)\nabla_{\mathbf{y}}\Phi_0(\mathbf{y})d^3y, 
\end{equation}
~~~~~~~~~~~~~~~~~~~~~~~~~~~............................
\begin{equation}\label{10}
 \nabla \Phi_n =  \nabla \Phi_N +  \boldsymbol{\Psi}_{n-1} - \int K(\mathbf{u}, v_{n-1})\nabla_{\mathbf{y}}\Phi_{n-1}(\mathbf{y})d^3y, 
\end{equation}
\begin{equation}\label{11}
 \nabla \Phi_{n+1} =  \nabla \Phi_N +  \boldsymbol{\Psi}_n - \int K(\mathbf{u}, v_n)\nabla_{\mathbf{y}}\Phi_n(\mathbf{y})d^3y, 
\end{equation}
and so on. Let us recall here that $\Bbbk(\mathbf{x},\mathbf{y})=K(\mathbf{u},v)$ and we have extended the definition of $v$ in Eq.~\eqref{3} such that $v_n=|\nabla_{\mathbf{y}}\Phi_n(\mathbf{y})| / |\nabla_{\mathbf{x}} \Phi_n(\mathbf{x})|$. Moreover, $\boldsymbol{\Psi}_0,  \boldsymbol{\Psi}_1, ...,  \boldsymbol{\Psi}_n, ...$ are such that the integrability condition is satisfied at each step of the approximation process, namely, 
\begin{equation}\label{12}
 \nabla \times  \boldsymbol{\Psi}_n=\int \nabla_{\mathbf{x}} K(\mathbf{u},v_n) \times \nabla_{\mathbf{y}}\Phi_n(\mathbf{y})d^3y
\end{equation}
for $n=0,1,2,...$. Here $\boldsymbol{\Psi}_n$, for instance, can be expressed in terms of $\Phi_n$ using the method described in Eqs.~\eqref{7} and \eqref{8}; then, the result may be employed in the expression for $\nabla \Phi_{n+1}$ in Eq.~\eqref{11} of the successive approximation scheme. As $n \to \infty$, we expect that $\Phi_n$ approaches $\Phi$ and $\boldsymbol{\Psi}_n$ approaches  $\boldsymbol{\Psi}$, so that the limiting form of Eq.~\eqref{12} coincides with Eq.~\eqref{6}. The \emph{convergence} of this successive approximation process depends of course upon the nature  of the kernel and its treatment is beyond the scope of this work, as the general form of kernel $K(\mathbf{u},v)$ is unknown at present. 

The general solution of Eq.~\eqref{2} presented in this section can be used, in principle, to restrict the form of kernel $K(\mathbf{u},v)$ on the basis of observational data. Nonlocal gravity simulates dark matter~\cite{nonlocal, NonLocal, BCHM, BM}; therefore, it may be possible to determine the general \emph{nonlinear} kernel $K$ from the comparison of our general solution of Eq.~\eqref{2} with astrophysical data regarding dark matter. However, the treatment of this general inverse problem of nonlocal gravity is a task for the future. 

\section{Gravitational Potential of a Point Mass}

As a simple application of the formal procedure developed in the previous section, we will consider here the gravitational potential due to a point mass $M$ at the origin of spatial coordinates, so that $\rho(\mathbf{x})=M\delta(\mathbf{x})$. The corresponding Newtonian potential is $\Phi_N=-GM/r$ and we expect that $\Phi$ is also just a function of $r$ as a consequence of the spherical symmetry of the point source. Similarly, it is natural to assume that the kernel's dependence on $\mathbf{u}$ is only through its magnitude $u$ due to the isotropy of the source.  A detailed investigation reveals that $\boldsymbol{\Psi}=0$ in this case; we outline below the main steps in this analysis. 

In computing the integral term in Eq.~\eqref{5}, we introduce the spherical polar coordinate system $(y, \vartheta, \varphi)$ in which the polar axis is taken to be along the $\mathbf{x}$ direction. The kernel in Eq.~\eqref{5} is then just a function of $r$, $y$  and $\cos\vartheta$; moreover, 
$\nabla_{\mathbf{y}}\Phi(\mathbf{y})$ equals $d\Phi(y)/dy$ times the unit vector in the $y$ direction. The azimuthal components of this unit vector vanish upon integration over all angles and only its polar component remains. Therefore, $\boldsymbol{\Psi}$ is purely radial in this case, namely,
\begin{equation}\label{13}
\boldsymbol{\Psi} = \chi(r) \mathbf{x}\,,
\end{equation}
which satisfies the integrability condition given in Eq.~\eqref{6}, since in this case the curl of $\boldsymbol{\Psi}$ identically vanishes. Here $\chi(r)$ can be determined from the requirement that $\nabla \cdot \boldsymbol{\Psi}=0$. It then follows that $\chi=m/r^3$, where $m$ is an integration constant. Thus $\boldsymbol{\Psi}=m\mathbf{x}/r^3$; that is, the right-hand side of Eq.~\eqref{5} is radial in direction and is given by $(GM+m)\mathbf{x}/r^3$. The resulting $\boldsymbol{\Psi}$ in effect indicates the presence of an extra delta-function source at the origin of spatial coordinates. We therefore set $m$, which is effectively a new mass parameter, equal to zero, as it simply renormalizes the mass of the source. Thus $\boldsymbol{\Psi}=0$ and Eq.~\eqref{5} reduces in this case to 
\begin{equation}\label{14a}
 \frac{d\Phi}{dr} + 2 \pi \int_0^{\pi} \int_0^{\infty}K(u, v) \frac{d\Phi(y)}{dy}y^2dy \cos \vartheta \sin \vartheta d\vartheta = \frac{GM}{r^2},
\end{equation}
where $u$ and $v$ are given by
\begin{equation}\label{14b}
u =  \sqrt{r^2+y^2-2ry\cos \vartheta} \,, \quad v = |\frac{d\Phi(y)}{dy}/\frac{d\Phi(r)}{dr}|\,.
\end{equation}
The extra factor of $\cos \vartheta$ in Eq.~\eqref{14a} is due to the fact that in Eq.~\eqref{5} the component of $\mathbf{y}/y$ along the polar axis is  $\cos \vartheta$.

\subsection{Linear Kernel}

Let us assume, for the sake of simplicity, that $\Bbbk(\mathbf{x},\mathbf{y})=k(u)$, so that in this subsection we are only concerned with a nonlocally  modified Poisson's equation that is \emph{linear} with a kernel that depends only on $u$ as a result of the spherical symmetry of the point source. Then, Eq.~\eqref{14a} for the \emph{linear} gravitational potential $\Phi_{\ell}$ reduces to
\begin{equation}\label{14}
 \frac{d\Phi_{\ell}}{dr} + 2 \pi \int_0^{\pi} \int_0^{\infty}k(\sqrt{r^2+y^2-2ry\cos \vartheta} \,)\frac{d\Phi_{\ell}(y)}{dy}y^2dy \cos \vartheta \sin \vartheta d\vartheta = \frac{GM}{r^2}\,,
\end{equation}
which means that a linear integral operator with kernel $k$ acting on $d\Phi_{\ell}/dr$ results in $GM/r^2$. We note in passing that the successive approximation method of the previous section  leads in this case to the standard Liouville-Neumann solution of Eq.~\eqref{14} via iterated kernels of the Fredholm integral equation of the second kind~\cite{T}; however, as discussed Sec. II, the Neumann series diverges in this case and the corresponding solution does not exist under physically reasonable conditions. Therefore, we adopt the Fourier transform method and let $q(u)$ be the kernel that is reciprocal to $k(u)$; then, $d\Phi_{\ell}/dr$ is given by the linear integral operator, with $k$ replaced by $q$, acting on $GM/r^2$. That is, 
\begin{equation}\label{15}
 \frac{d\Phi_{\ell}}{dr} = \frac{GM}{r^2} + 2 \pi GM \int_0^{\pi} \int_0^{\infty}q(\sqrt{r^2+y^2-2ry\cos \vartheta}  \,)dy \cos \vartheta \sin \vartheta d\vartheta \,.
\end{equation}

Substituting the Kuhn kernel~\eqref{2i} for $q$ in Eq.~\eqref{15} and performing the $y$-integration in the resulting integral first, we find that for $\vartheta \in (0, \pi]$,
\begin{equation}\label{19}
\int_0^{\infty}\frac{dy}{(y-r\cos \vartheta)^2+r^2\sin^2 \vartheta}=\frac{\pi - \vartheta}{r \sin \vartheta}\,.\end{equation}
The $\vartheta$-integration is then straightforward and the end result is
\begin{equation}\label{20}
\frac{d\Phi_{\ell}}{dr}=\frac{GM}{r^2} + \frac{GM}{\lambda} \frac{1}{r}\,,
\end{equation}
in agreement with the radial derivative of Eq.~\eqref{2g}. In this way, starting from our general solution of the modified Poisson equation, we again recover  the Tohline-Kuhn scheme of modified gravity.

\subsection{Nonlinear Kernel}

To gain some insight into the  role of  nonlinearity in Eq.~\eqref{14a}, let us suppose that nonlinearity constitutes a very small perturbation on a background linear kernel. In fact, we set $K(u,v) = k(u) + \epsilon P(u, v_{\ell})$, where $\epsilon$, $0<\epsilon \ll 1$, is a sufficiently small parameter and $v_{\ell}$ is obtained from $v$ in Eq.~\eqref{14b} by replacing $\Phi$ with $\Phi_{\ell}$. We thus expand $\Phi$ to first order in $\epsilon$ and thereby develop a simple linear perturbation theory for Eq.~\eqref{14a} such that
\begin{equation}\label{21}
\Phi =\,\Phi_{\ell} + \epsilon \Phi_{n\ell}\,.
\end{equation}
Here $\Phi_{\ell}(r)$ is given in general by Eq.~\eqref{15} and $\Phi_{n\ell}$ is the perturbation potential due to nonlinearity. Moreover, Eq.~\eqref{14a} implies that 
\begin{equation}\label{22}
 \frac{d\Phi_{n\ell}}{dr} + 2 \pi \int_0^{\pi} \int_0^{\infty}k(\sqrt{r^2+y^2-2ry\cos \vartheta} \,)\frac{d\Phi_{n\ell}(y)}{dy}y^2dy \cos \vartheta \sin \vartheta d\vartheta = N(r)\,,
\end{equation}
where $N(r)$ is due to the nonlinear part of the kernel and is given by
\begin{equation}\label{23}
N(r)=- 2 \pi \int_0^{\pi} \int_0^{\infty} P(u, v_{\ell})\frac{d\Phi_{\ell}(y)}{dy}y^2dy \cos \vartheta \sin \vartheta d\vartheta \,.
\end{equation}
As in the previous subsection, Eq.~\eqref{22} can be solved by means of kernel $q(u)$ that is reciprocal to $k(u)$ and we find
\begin{equation}\label{24}
 \frac{d\Phi_{n\ell}}{dr} = N(r) + 2 \pi \int_0^{\pi} \int_0^{\infty}q(\sqrt{r^2+y^2-2ry\cos \vartheta}  \,)N(y)y^2 dy \cos \vartheta \sin \vartheta d\vartheta \,.
\end{equation}
A consequence of this result should be noted here: Inspection of Eqs.~\eqref{15}, \eqref{23} and \eqref{24} reveals that $\Phi_{n\ell}(r)$ is simply proportional to the gravitational constant $G$. This feature is an example of the general scaling property of Eq.~\eqref{2}, which implies that any solution $\Phi$ of Eq.~\eqref{2} must be proportional to $G$. 

It is therefore possible to see that our nonlocal as well as nonlinear modification of Newtonian gravity cannot behave as in the Modified Newtonian Dynamics (MOND) approach to the breakdown of Newtonian gravity~\cite{M, SM, Ibata}.  From the scaling property of our modified Poisson's equation, we expect that the gravitational potential $\Phi$ is in general proportional to the gravitational constant $G$, since the source term in Eq.~\eqref{2} is proportional to $G$. Therefore, the nonlocal theory, as a consequence of its particular \emph{nonlinear} form in the Newtonian domain, does \emph{not} contain a MOND regime, where the gravitational potential would then be proportional to $G^{1/2}$.

\section{Discussion}

In recent papers~\cite{nonlocal, NonLocal, BCHM, BM}, nonlocality has been introduced into classical gravitation theory via a scalar kernel $k$. However, observational data can provide information about its reciprocal kernel $q$. This is similar to the situation in general relativity, where gravitation is identified with spacetime curvature, but observations generally do not directly measure the curvature of spacetime, except possibly in relativistic gravity gradiometry. We make a beginning in this paper in the treatment of the inverse problem of nonlocal gravity. The scalar nonlocal kernel $k$ must be determined from observational data that involve the reciprocal kernel $q$. Our preliminary study involves the Newtonian regime, where the nonlocally modified Poisson's equation is investigated in its linearized convolution form. We present a detailed mathematical analysis of the resulting Fredholm integral equation using the Fourier transform method and prove the existence of the nonlocal convolution kernel $k(\mathbf{u})$ when its reciprocal $q(\mathbf{u})$ satisfies certain physically reasonable conditions. Simple explicit examples are worked out in connection with the linear gravitational potential of spiral galaxies. To extend our treatment beyond the Newtonian domain, it would be necessary to consider relativistic generalizations of the Kuhn kernel along the lines indicated in Sec. III of Ref.~\cite{BCHM}.

Next, we present a general treatment of the nonlocal and \emph{nonlinear} modification of Poisson's equation that represents nonlocal gravity in the Newtonian regime. The method of successive approximations is then employed to provide a formal solution.  The utility of this general approach is illustrated for the determination of the gravitational potential of a point mass when nonlinearities are assumed to be relatively small. In this case, we recover anew the Tohline-Kuhn phenomenological modified gravity approach to the dark matter problem in astrophysics~\cite{Tohline,Kuhn,Jacob1988}. 

To place our work in the proper context, we note that nonlocal special relativity (developed since 1993, cf.~\cite{BM2}) and the principle of
equivalence imply the necessity of a nonlocal generalization of Einstein's
theory of gravitation. Here nonlocality is encoded in a nonlocal
``constitutive" kernel $k$ that must be determined from observation. In
working out the physical consequences of nonlocal gravity, it was soon
discovered~\cite{nonlocal, NonLocal} that it reproduces the 1980s Tohline-Kuhn phenomenological
approach to dark matter as modified gravity~\cite{Tohline, Kuhn}. This connection is the most
fundamental contact of the new theory with observation and indicates to us
that we are on the right physical track. To verify this, we must compute the
nonlocal kernel $k$ from the rotation curves of spiral galaxies and show that
it has the proper physical properties expected of such a kernel. Our present
paper accomplishes this task. That is, we extend the Kuhn kernel $q$ analytically to all space and then use the
result to solve the inverse problem of finding kernel $k$ by means of Fourier
integral transforms. The resulting $k$ has indeed just the expected properties
and puts the nonlocal theory of gravity on a more solid observational foundation.

\begin{acknowledgments}
B.M. is grateful to F.W. Hehl and J.R. Kuhn for valuable comments and helpful correspondence. 
\end{acknowledgments}

\appendix{}
\section{Radial Convolution Kernels}\label{app}

The purpose of this appendix is to present some useful relations between the radial convolution kernels that we employ in Sec. III. We use dimensionless quantities throughout. 

A radial convolution kernel $q(u)$ can be expressed in terms of its Fourier transform $\hat{q}(\xi)$ as
\begin{equation}\label{A1}
q(u)=\frac{1}{2\pi^2 u}\int_0^{\infty}\xi \hat{q}(\xi)\sin(\xi u)d\xi\,. 
\end{equation}
Eqs.~\eqref{A1} and~\eqref{B2} form a pair of Fourier sine transforms such that
\begin{equation}\label{A2}
q(0)=\frac{1}{2\pi^2}\int_0^{\infty}\xi^2 \hat{q}(\xi)d\xi\,, 
\end{equation}
\begin{equation}\label{A3}
\hat{q}(0)=4\pi \int_0^{\infty}r^2 q(r)dr\,. 
\end{equation}
It is interesting to note that Eqs.~\eqref{B5} and~\eqref{A2} imply that
\begin{equation}\label{A4}
k(0)=-\frac{1}{2\pi^2}\int_0^{\infty}\frac{\xi^2 \hat{q}(\xi)}{1+\hat{q}(\xi)}d\xi\,,
\end{equation}
\begin{equation}\label{A5}
k(0)+q(0)=\frac{1}{2\pi^2}\int_0^{\infty}\frac{\xi^2 \hat{q}^2(\xi)}{1+\hat{q}(\xi)}d\xi\,.
\end{equation}

It is clear from a comparison of Eqs.~\eqref{A3} and~\eqref{B19} that $\hat{q}(0)\approx M_D/M$. Therefore, we can conclude from the discussion in Sec. III that $\hat{q}(0)\approx 2/\alpha$ in Figures 1 and 3, in agreement with our numerical results. 

For the two particular examples considered in Sec. III, $q(0)=(1+a\alpha)/(4\pi a^2)$ for the first example given by Eqs.~\eqref{BB1} and~\eqref{BB2}, and $q(0)=\infty$ for the second example that has a central cusp and is given by Eq.~\eqref{C1}.  Let us note that in either case $\hat{q}(\xi)$ is square integrable over the whole $\boldsymbol{\xi}$ space, so that $\xi^2\hat{q}^2(\xi)$ is integrable over the radial coordinate $\xi: 0\to \infty$; therefore, the right-hand side of Eq.~\eqref{A5} is finite. It then follows from Eq.~\eqref{A5} that  $k(0)$ is finite in the first example due to the finiteness of $q(0)$, while in the second example, $q(0)=\infty$ and hence $-k(0)=\infty$, in agreement with the numerical results of Figures 2 and 4.

Finally, let $q'(u)$ and $q''(u)$ represent respectively the reciprocal kernels given in Sec. III in the first and second examples; then,  
\begin{equation}\label{A6}
q''(u)-q'(u)=\frac{a}{4\pi} \frac{1+\alpha(a+u)}{u(a+u)^2} e^{-\alpha u}\,. 
\end{equation}
Moreover, we find from Eqs.~\eqref{A6} and~\eqref{B2} that 
\begin{equation}\label{A7}
\hat{q}''(\xi)-\hat{q}'(\xi)=\frac{a}{\xi}\int_0^{\infty}\Big[\frac{1}{(a+r)^2}+\frac{\alpha}{a+r}\Big]e^{-\alpha r}\sin(\xi r)~dr\,. 
\end{equation}
It then follows from the lemma given at the end of Sec. II that the right-hand side of Eq.~\eqref{A7} is positive. Thus for any $\xi \ge 0$, $\hat{q}''(\xi) > \hat{q}'(\xi)$; moreover, as $\xi \to \infty$, $\hat{q}''(\xi)-\hat{q}'(\xi)\to 0$, in accordance with the Riemann-Lebesgue lemma.

\end{document}